# Design of a Simplified Wireless Sensor Network Node based on MQTT Protocol

Zhen-tang SHAO[1,2], Meng-xing HUANG[1,2,*], Di WU[1,2], Xu ZHANG[1,2] and Ao HUANG[1,2]

[1]State Key Laboratory of Marine Resource Utilization in South China Sea, Haikou, 570228, China

[2]College of Information Science and Technology, Hainan University, Haikou, 570228, China

* Corresponding author e-mail:huangmx09@163.com

**Abstract.** MQTT protocol is a publish/subscribe message protocol based on TCP/IP protocol. It has the characteristics of low power consumption, scalability, openness and simplicity. HTTP protocol is an open and low-cost request/reply message protocol based on TCP/IP protocol. It is the main protocol of Internet communication, but it is not suitable for the environment of computing, processing and bandwidth limited. In this paper, a simplified wireless sensor network node was designed based on MQTT protocol. The node was designed to use the Arduino development environment and use the WiFi for networking. It has the characteristics of simple structure, low power consumption and so on. The design can be widely used in smart home, environmental monitoring and medical applications, and is the main contents of the Internet of things (IoT).

## Introduction

Due to the popularity of IoT technology, its main application of wireless sensor network technology also shows broad application prospects. Wireless sensor networks cooperate with each other to collect, transmit and interact data through the deployment of sensors [1]. The wireless sensor network (WSN) forms a self-organizing network system through a large number of nodes. The node is the basic component of WSN. Its energy consumption, stability and transmission speed are related to the work efficiency and work cost of the whole network. Therefore, designing a simple and low cost wireless sensor node is the key to the design of wireless sensor network [2].

The IoT consists of three levels, namely, the perception layer, the network layer and the application layer. The communication protocol of the Internet of things is mainly responsible for communication and data exchange between devices. Nowadays, the mainstream protocol is the HTTP protocol, which has high openness, good operability and low cost, so it is widely used in the Internet of things. But in the Internet of things, small micro embedded devices have the characteristics of low bandwidth and high delay. Because of the high protocol overhead, HTTP is not suitable for resource shortage embedded systems [3]. Compared with HTTP, MQTT has the characteristics of low power consumption, openness, simplicity and easy operation [4]. It is suitable for the environment with limited resources. In this paper, we design and verify the wireless sensor network node based on MQTT protocol.

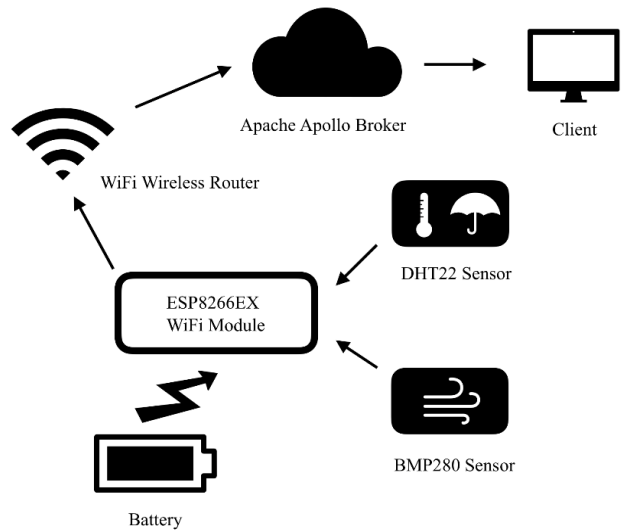

Figure 1. System configuration

Fig. 1 shows the design structure of the whole system. The whole system uses the ESP8266EX chip as the core controller to read the sensor data and transmit it to the server. The ESP8266EX chip is connected with the DHT22 temperature and humidity sensor and the BMP280 air pressure sensor. At the same time, the ESP8266EX chip is connected to the local WiFi wireless router, and the acquired sensor data is uploaded to the Apache Apollo server based on the Linux. The server opens the standard RESTful API for the external devices to read the sensor data easily and conveniently at anytime and anywhere. At the same time, using ECharts to display data, you can easily see the current environment by the web pages.

The system uses 9.62Wh capacity of 18650 Li-ion battery for system power supply, and uses the TC4056A chip to control the battery charging. Besides, the Micro USB interface is added. The core control unit of the system is developed with Arduino IDE, and the related programs are written. The system uses WiFi for data transmission. Compared with Zigbee and Bluetooth, the system does not need additional gateways. It can be more convenient to access the existing network and save the cost of building a sensor network. At the same time, thanks to the excellent characteristics of MQTT protocol, which can accept 1000000 nodes concurrently, and is suitable for large-scale space environment information detection. The use of rechargeable Li-ion battery design can reduce the cost of battery charges, better protect the environment and reduce the environmental pollution of waste batteries.

**Literature Survey**

The MQTT protocol is a protocol designed for communication with remote sensors and control devices with limited computing power and unreliable network [5]. In the application of Internet of things, MQTT protocol is very popular. It has the following characteristics [6]:
1. Use publish / subscribe message mode to provide one to many message releases and release application coupling;
2. Using TCP/IP to provide network connections[7];
3. The overhead is very small (the fixed length of the head is 2 bytes), small transmission, minimization of protocol exchange to reduce network traffic
4. Use Last Will and Testament features to notify the parties concerned about the mechanism of client's abnormal interruption;
5. Transmission of messages shielded by the content of the load

The system uses the MQTT 3.1.1 version of the OASIS standard, uploading data through the Public control message in the protocol, and setting the quality of service to 0 (QoS 0). When the data upload error occurs, do not reupload, in order to save resources and bandwidth.

**Circuit and Working**

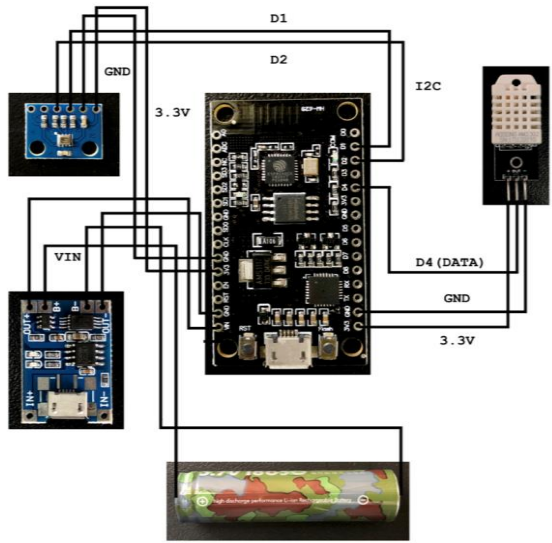
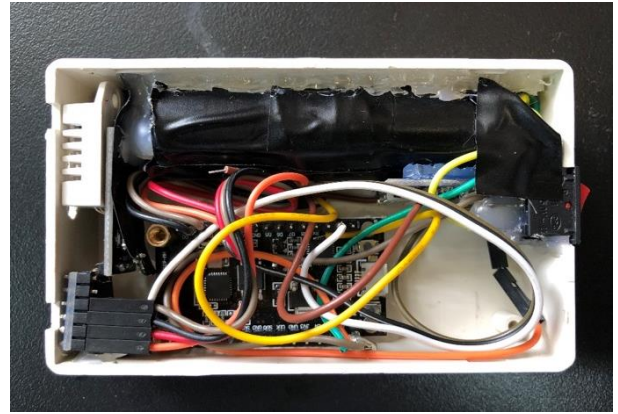

Figure 2. Schematic diagram        Figure 3. The internal structure of the sensor node

Fig. 2 is a circuit design. In the design of the circuit, the ESP8266EX uses a 18650 Li-ion battery for power supply, at the same time the Li-ion battery drives the DHT22 sensor and the BMP280 sensor. The expansion board provides 11 GPIO ports and power input and output interface. In addition, the front end of ESP8266EX also adds a conversion chip named CP2102 between USB and TTL serial ports to facilitate the programming of the connection between the chip and the computer. A GPIO serial port connected LED indicator is also added, so that we can check the correctness of the program and the communication status of the serial port. Fig. 3 is the internal structure of the sensor node.

**Hardware Aspect**

**ESP8266EX.**

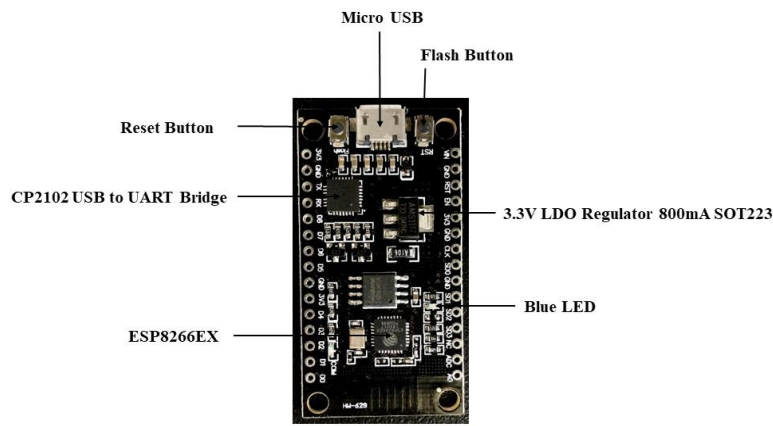

Figure 4. ESP8266EX

ESP8266EX is an ultra low power and high integration Wi-Fi chip [8]. ESP8266EX has Tensilica L106 32 bit RISC processor, CPU clock speed up to 160 MHz, supports real-time operating system (RTOS) and Wi-Fi protocol stack [9]. It is designed for mobile devices, wearable electronic products and Internet of things, and has advanced low power management technology [10]. The working temperature range of ESP8266EX is large, and it can maintain stable performance, and can adapt to various operating environments [11]. In addition, ESP8266EX can connect sensors and other devices through GPIO.

Table 1. ESP8266EX specification

| Specification Parameters | Values |
|---|---|
| MCU | 32bit TenSilica L 106 |
| Clock Speed | 80MHz / 160 MHz |
| RAM | <36Kb |
| Operating Voltage | 3.0V – 3.6V |
| Operating Current | 80mA (Average) |
| Available GPIO pins | 10 |

**DHT22.**

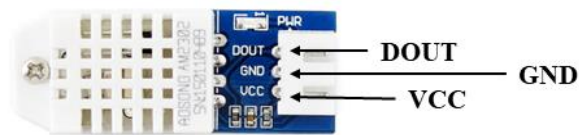

Figure 5. DHT22

Fig. 5 is a physical map of DHT22. DHT22 is a calibrated digital temperature and humidity sensor for detecting environmental temperature and humidity [12]. It uses digital signal technology to ensure reliability and stability while consuming less power and long-term stability. Fig. 6 is the external connection diagram of DHT22. In Table 2, we compare the three sensors, DHT22, DHT11 and SHT31, in terms of cost and specifications.

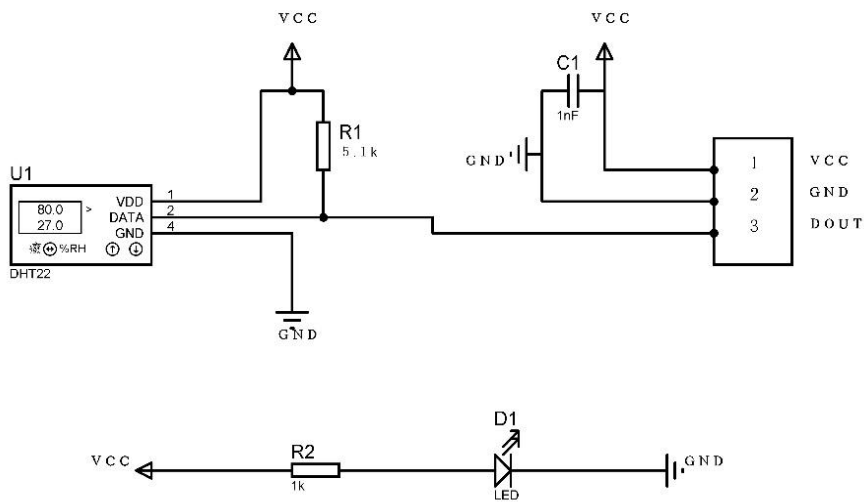

Figure 6. DHT22 external connection diagram

Table 2. Compare DHT22, DHT11 and SHT31

| Parameter | DHT22 | DHT11 | SHT31 |
|---|---|---|---|
| Humidity Range | 0 - 100% | 20 - 80% | 0 - 100% |
| Temperature Range | -40-80°C | 0-50°C | -40-90°C |
| Accuracy | ±2% (Humd) ±0.5°C (Temp) | ±5% (Humd) ±2°C (Temp) | ±2% (Humd) ±0.3°C (Temp) |
| Repeatability | ±0.3% (Humd) ±0.2°C (Temp) | ±1% (Humd) ±1°C (Temp) | ±0.1% (Humd) ±0.06°C (Temp) |
| Typical Price | $4-10 | $1-5 | $8-15 |

**BMP280.**

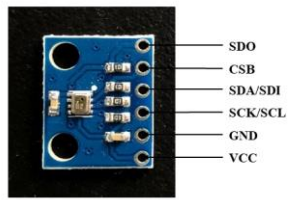

Figure 7. BMP280

Fig. 7 is BMP280, a pneumatic sensor designed for mobile applications. The sensor module is extremely compact and has high accuracy and flexibility. Because of its small size and low power consumption, it is allowed to be implemented in electronic devices such as mobile phones, GPS modules and watches.

**Software Aspect**

**Arduino IDE.** Arduino IDE is a software used to develop Arduino programs. After writing the program, it can be uploaded to the Arduino development board or other compatible development boards through this software [13]. The Arduino IDE used in this paper has added the supporting files of ESP8266 series chip, and programmed the program through serial port. By function, the whole program realizes the encrypted connection of WiFi router, the reading of sensor data, the construction of JSON format text, and the landing of MQTT server and the data uploading. From the Fig. 8, we can see that the serial port monitor displays the data read and uploaded successfully.

Figure 8. Serial port monitor displays data uploading and uploading successfully

**MQTT Server.** The entire system server uses the Apache Apollo broker under the Linux system to support a variety of protocols such as STOMP, AMQP, MQTT, Openwire, SSL, and WebSockets. MQTT is a succinct binary protocol that is suitable for such resource constrained and unstable network conditions. Apollo allows clients to connect through the open MQTT protocol. The protocol is mainly driven by limited resources and unstable network. It is a subscription and publish model. Any client who implements MQTT can connect to Apollo. We can view information uploaded by sensor nodes through the management page of Apollo. All uploaded information data is stored in the LevelDB format database of the server, and the data in the database can be invoked by programming.

**ECharts.** The system results are displayed using the ECharts framework. ECharts is an open source visual library implemented by JavaScript. It can run smoothly on PC and mobile devices, and is compatible with most browsers. We call the RESTful API interface of the MQTT server through the ECharts framework, get the data once every ten seconds, parse the files obtained in the JSON format, and get the line map of the current sensor data to display through the online web page.

## Experimental Result

We demonstrate the design of sensor node system through data presentation, as shown in Fig. 9. When the switch of the wireless sensor node is power on, it will automatically connect to our WiFi router. When the connection is successful, the node will try to log on to the MQTT server and confirm whether the connection is successful. After the connection is successful, the ESP8266EX chip will acquire the sensor data and construct the JSON format to upload.

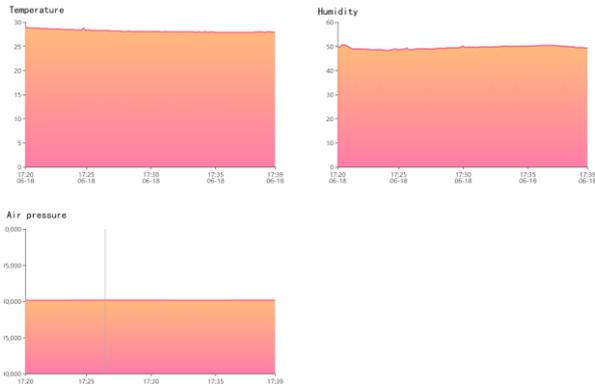

Figure 9. The sensor data displaied on an online web page

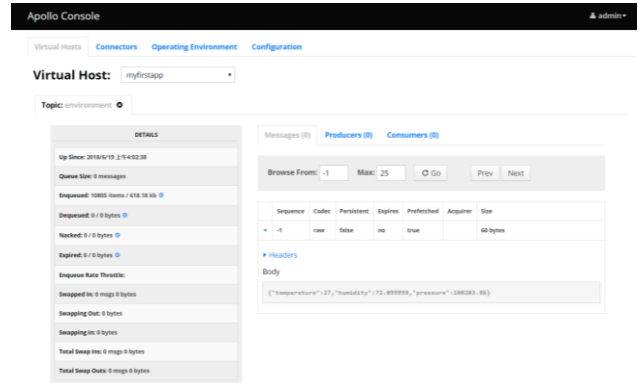

Figure 10. Apollo console displays the sensor data

We demonstrate the effectiveness of our entire design through the Apollo console and ECharts online icon, as shown in Fig. 10. At the same time, we get the data of RESTful API through the Python script, as shown in Fig. 11.

```
inner humidity: 72.099998%
inner pressure: 100203.86 Pa
inner temperature: 27*C
inner humidity: 72.099998%
inner pressure: 100203.86 Pa
inner temperature: 27*C
inner humidity: 72.099998%
inner pressure: 100203.86 Pa
inner temperature: 27*C
inner humidity: 72.099998%
inner pressure: 100203.86 Pa
inner temperature: 27*C
inner humidity: 72.099998%
inner pressure: 100203.86 Pa
inner temperature: 27*C
inner humidity: 72.099998%
inner pressure: 100203.86 Pa
inner temperature: 27*C
inner humidity: 72.099998%
inner pressure: 100203.86 Pa
inner temperature: 27*C
inner humidity: 72.099998%
inner pressure: 100203.86 Pa
inner temperature: 27*C
```

Figure 11. The Python script gets the data of the RESTful API

## Conclusion

In this paper, we have realized the design of wireless sensor node system based on MQTT protocol through experiments. The system can successfully upload data through nodes to the server and easily access standard JSON format data through open RESTful API to facilitate developers to invoke these data and carry out a variety of applications.

The design of this system can help smart home and smart city system develop and put into use more quickly.

## Acknowledgment

This work was supported by the National Natural Science Foundation of China (Grant #: 61462022), the National Key Technology Support Program (Grant #: 2015BAH55F04, Grant

#:2015BAH55F01), Major Science and Technology Project of Hainan province (Grant #: ZDKJ2016015), Natural Science Foundation of Hainan province (Grant #: 614232 and Grant#:617062), Scientific Research Staring Foundation of Hainan University (Grant #: kyqd1610).